\documentclass[prd,altaffilletter,twocolumn,superscriptaddress,floatfix,nofootinbib]{revtex4} 

\newif\iflonger
\longertrue

\newif\ifsection
\sectionfalse

\usepackage{amsfonts,amsmath,graphicx,bm,color,dcolumn}

\newcolumntype{d}[1]{D{.}{.}{#1} }

\usepackage[a-1b]{pdfx}
\hypersetup{colorlinks=true,
  urlcolor=blue,       
  filecolor=green,     
  linkcolor=red}

\newcommand{\cambridge}{DAMTP,
University of Cambridge, Cambridge CB3 0WA,
  United Kingdom}

\newcommand{\cornell}{Laboratory of Elementary Particle Physics,
  Cornell University, Ithaca, NY 14853, United States}

\newcommand{\glasgow}{SUPA, School of Physics and Astronomy,
  University of Glasgow, Glasgow, G12 8QQ, United Kingdom}

\newcommand{\instnuclth}{Institute for Nuclear Theory,
  University of Washington, Seattle, WA 98195-1550, United States}

\newcommand{\ohiostate}{Department of Physics, Ohio State University,
  Columbus, OH 43210, United States}
  
\newcommand{\wm}{Physics Department, College of William and Mary, Williamsburg, VA 23187, 
United States}

\newcommand{\jlab}{Thomas Jefferson National Accelerator Facility, 
Newport News, VA 23606, United States}

\newcommand{\VEC}[1]{{\bf \bm{#1}}} 
\newcommand{\subrm}[1]{{\scriptscriptstyle\mathrm{#1}}}
\newcommand{\order}[1]{ {\mathcal{O}(#1)} } 
\newcommand\Dl{\raisebox{0.05em}{$\stackrel{\scriptstyle\leftarrow}D$}}

\def\today{\number\day\space\ifcase\month\or
January\or February\or March\or April\or May\or June\or
July\or August\or September\or October\or November\or December\fi
\space\number\year}
\newcount\mins \newcount\hours
\def\now{\hours=\time \mins=\time
	\divide\hours by60 \multiply\hours by60 \advance\mins by-\hours
	\divide\hours by60 
	\number\hours:\ifnum\mins<10 0\fi\number\mins }
	
\usepackage{color}

\begin{document}

\title{
Lattice QCD matrix elements for the $\bm{B_s^0 - \bar{B}_s^0}$ width difference beyond leading order
}

\author{Christine~T.~H.~\surname{Davies}} 
\affiliation{\glasgow}

\author{Judd \surname{Harrison}} 
\affiliation{\cambridge}
\affiliation{\glasgow}

\author{G.~Peter~\surname{Lepage}}
\affiliation{\cornell}

\author{Christopher~J.~\surname{Monahan}}
\affiliation{\instnuclth}
\affiliation{\wm}
\affiliation{\jlab}

\author{Junko~\surname{Shigemitsu}}
\affiliation{\ohiostate}

\author{Matthew \surname{Wingate}}
\email[]{M.Wingate@damtp.cam.ac.uk}
\affiliation{\cambridge}

\collaboration{HPQCD Collaboration}
\email[]{http://www.physics.gla.ac.uk/HPQCD}


\begin{abstract}
Predicting the $B_s^0-\bar{B}_s^0$ width difference $\Delta\Gamma_s$ relies on the heavy quark expansion and on hadronic matrix elements of $\Delta B=2$ operators.  We present the first lattice QCD results for matrix elements of the dimension-7 operators $R_{2,3}$ and linear combinations $\tilde{R}_{2,3}$ using nonrelativistic QCD for the bottom quark and a highly improved staggered quark (HISQ) action for the strange quark.  Computations use MILC ensembles of gauge field configuations with $2+1+1$ flavors of sea quarks with the HISQ discretization, including lattices with physically light up/down quark masses.  We discuss features unique to calculating matrix elements of these operators and analyze uncertainties from series truncation, discretization, and quark mass dependence.  Finally we report the first Standard Model determination of $\Delta\Gamma_s$ using lattice QCD results for all hadronic matrix elements through $\mathcal{O}(1/m_b)$.  The main result of our calculations yields the $1/m_b$ contribution $\Delta \Gamma_{1/m_b} = -0.022(10)~\mathrm{ps}^{-1}$. Adding this to the leading order contribution, the Standard Model prediction is $\Delta \Gamma_s = 0.092(14)~\mathrm{ps}^{-1}$.
  
\end{abstract}

\maketitle

\ifsection
\section{Introduction}
\fi

Oscillations of neutral mesons into their antiparticles have been important phenomena in the study of quark flavor. This flavor-changing, neutral mixing is absent in the Standard Model (SM) at the classical level; appearing at one-loop level it is suppressed by two powers of Fermi's constant $G_F$ relative to hadronic and quark mass scales. A few observables are, to a high level of precision, sensitive only to short-distance physics. Within the Standard Model, predictions for these are reliably calculable because the dominant contribution comes from top-quark loops; there is no significant contribution from intermediate-state hadronic physics.  Prime examples are:  in the mixing of strange mesons $K^0-\bar{K}^0$, the indirect CP-violating ratio $\epsilon_K$ and,  in beauty mesons $B^{0}_{d,s} - \bar{B}^0_{d,s}$, the mass differences (equivalently, the oscillation frequencies) $\Delta M_{d,s}$.   Precise experimental measurements of these, together with accurate Standard Model predictions, constitute stringent tests of the SM description of quark flavor.

Beyond these observables are others, where contributions from hadronic intermediate states must be included.  For example, mixing in neutral charm mesons $D^0-\bar{D}^0$ has significant long-distance contributions due to differing flavor structure from $K^0$ and $B^0$ mixing.  Predictions for heavy meson and baryon lifetimes also require theoretical treatment of long-distance effects.

The $B^{0}_{s} - \bar{B}^0_{s}$ width difference $\Delta \Gamma_s$ is another example and is the focus of this paper.  Unlike the mass difference, which comes from the real part of the mixing amplitude, the width difference comes from the imaginary part which, by the optical theorem, describes the decays to real final states, primarily $b \to c\bar{c}s$ decays.   Consequently, $\Delta\Gamma_s$ is dominantly due to processes with intrinsic charm and we expect it to be insensitive to new physics. Comparison between theory and experiment is a test of the theoretical methods involved. Agreement here is a necessary condition for trusting these methods to reliably yield a SM prediction for quantities where new physics could contribute more prominently, e.g.\ in $D^0-\bar{D}^0$ mixing. 

Predicting the SM width difference $\Delta \Gamma_{s}$ requires the determination of matrix elements of a nonlocal product of effective operators ${H}_{\mathrm{eff}}^{\Delta F = 1}$, with charm and up quarks in the virtual loops.  Direct calculation of $\langle B^0_{s} |\mathcal{T}\{ {H}_{\mathrm{eff}}^{\Delta F = 1}(x) {H}_{\mathrm{eff}}^{\Delta F = 1}(0)\}| \bar{B}^0_{s}\rangle$ using lattice QCD is not presently feasible.  Therefore, an additional theoretical approximation is necessary in order to obtain a Standard Model prediction for $\Delta \Gamma_{s}$, namely the heavy quark expansion (HQE).  This expansion makes use of the large $b$-quark mass compared to the $c$-quark and other scales in the problem such as $\Lambda_{\subrm{QCD}}$, and approximates the imaginary part of the matrix elements above by a power series in $1/m_b$, composed of matrix elements of local, $\Delta F=2$ operators such as those appearing in ${H}_{\mathrm{eff}}^{\Delta F = 2}$ \cite{Beneke:1996gn}.
(See Ref.~\cite{Artuso:2015swg} for a recent review of the HQE applied to $\Delta\Gamma_s$.)

Matrix elements of the leading, dimension-6 operators in ${H}_{\mathrm{eff}}^{\Delta F = 2}$ have been calculated using lattice QCD with increasing precision, motivated by their impact on predictions for $\Delta M_{d,s}$.  Results are now available from several groups \cite{Carrasco:2013zta,Aoki:2014nga,Bazavov:2016nty,Dowdall:2019bea} (see Ref.~\cite{Aoki:2019cca} for a review and Refs.~\cite{Kirk:2017juj,King:2019lal} for sum rule calculations).  The precision of these determinations has become good enough that matrix elements of higher-dimension operators are needed in order to reduce the SM uncertainty in $\Delta\Gamma_s$.  Current estimates for the higher-dimension matrix elements come from the vacuum saturation approximation \cite{Artuso:2015swg}, although sum rules have also been applied \cite{Mannel:2007am}. The lack of any full QCD calculation of dimension-7 operators is the leading uncertainty in the Standard Model determination of $\Delta \Gamma_s$ \cite{Lenz:2011ti,Artuso:2015swg}.

In this paper, we provide results of the first complete lattice QCD calculation needed for $\Delta\Gamma_s$ through $O(1/m_b)$.  The goal here is to replace order-of-magnitude estimates based on the vacuum saturation approximation with first principles calculations including a quantitative analysis of errors.  In this first step, we neglect $\order{\alpha_s}$ corrections to the dimension-7 operators.  Including these corrections involves a technically challenging perturbative calculation which is left as a future project.

The convention we use for the dimension-6 operators $O_{1-5}$ is as in Ref.~\cite{Dowdall:2019bea}.
At higher order in the HQE, one needs matrix elements of the following operators
\begin{align}
  R_0 &= O_2 + \alpha_1 O_3 + \frac12 \alpha_2 O_1 \,, ~~~~~ R_1 = \frac{m_s}{m_b} O_4
   \nonumber \\
  R_2 & = \frac{1}{m_b^2} (\bar{b}^\alpha \Dl_\rho \gamma^\mu(1-\gamma^5)
  D^\rho s^\alpha)
  (\bar{b}^\beta \gamma_\mu(1-\gamma^5)s^\beta) \nonumber \\
  R_3 & = \frac{1}{m_b^2} (\bar{b}^\alpha \Dl_\rho (1-\gamma^5) D^\rho s^\alpha)
  (\bar{b}^\beta (1-\gamma^5)s^\beta) \
  \label{eq:Rops}
\end{align}
as well as the color-rearranged partners $\tilde{R}_1$, $\tilde{R}_2$, and $\tilde{R}_3$. The perturbative coefficients $\alpha_1$ and $\alpha_2$ are given in Refs.~\cite{Beneke:1998sy,Lenz:2006hd,Asatrian:2017qaz}.  
Using Fierz identities and neglecting terms at higher order in $1/m_b$ we have
\begin{equation}
\tilde{R}_2  = -R_2 ~~~\mbox{and}~~~\tilde{R}_3  = R_3 + \frac{1}{2}R_2 \,.
\label{eq:Rtilde}
\end{equation}
In nonrelativistic lattice QCD, matrix elements of both sides of (\ref{eq:Rtilde}) hold exactly.

\ifsection
\section{Description of lattice calculation}
\fi

We carry out our calculations using gauge field configurations generated by the MILC Collaboration \cite{Bazavov:2010ru,Bazavov:2012xda,Bazavov:2015yea}.  These  include the effects of $2+1+1$ flavors of sea quarks using the HISQ fermion action \cite{Follana:2006rc,Hart:2008sq}.  We use five separate ensembles, those labeled sets 1, 3, 4, 6, and 7 in \cite{Dowdall:2019bea}. 
\iflonger
(Also see the Appendix.)
\fi
Two of the ensembles have all of the quark masses tuned to be close to their physical values, e.g.\ $m_s/m_l\approx 27$;  these have lattice spacings of 0.12 and 0.15 fm.  The other three ensembles span 3 lattice spacings from 0.09 to 0.15 fm, with unphysically large light-quark masses corresponding to pion masses of about 300 MeV, $m_s/m_l\approx 5$.   Our use of three lattice spacings allows us to estimate discretization errors, and the computations done with unphysical light quark masses gives us information with which to correct any slight quark mass mistunings.  

The lattice actions used are the same as in our recent study of the dimension-6 operator matrix elements \cite{Dowdall:2019bea}.  Correlation functions are computed using the HISQ action for the strange quark; the valence quark mass is tuned to be closer to the physical strange mass than the value which was used for the sea strange quark.  The nonrelativistic QCD (NRQCD) action \cite{Lepage:1992tx,Dowdall:2011wh} is used for the bottom quark.  The action coefficients and valence quark parameters are the same as in recent work \cite{Dowdall:2011wh,Dowdall:2013tga,Dowdall:2019bea}. Because the determination of $\langle R_2 \rangle$ and $\langle R_3 \rangle$ here will have an $\order{\alpha_s}$ uncertainty due to tree-level matching between lattice and continuum regularization schemes, we only need a fraction of the statistics used in Ref.~\cite{Dowdall:2019bea}.  We will occasionally refer to Ref.~\cite{Dowdall:2019bea} as the high-statistics companion to this work.  Throughout this paper we will use the abbreviated notation $\langle \cdot \rangle \equiv \langle B_s | \cdot |\bar{B}_s\rangle$.

Let us examine a unique feature of computing $\langle R_2 \rangle$ and $\langle R_3 \rangle$. In the rest frame of the heavy quark, only the temporal component of the following bilinear is important at $1/m_b$ order
\begin{equation}
\frac{1}{m_b^2} (\bar{b}^\alpha \Dl_\rho \Gamma D^\rho s^\alpha) = 
\pm\frac{1}{m_b} (\bar{b}^\alpha \Gamma D^0 s^\alpha)  + \mathcal{O}\!\left(\frac{1}{m_b^2}\right)
\end{equation}
where the sign is determined by whether the temporal derivative acts on an outgoing heavy quark or an incoming heavy antiquark.  $\Gamma$ represents either $\gamma^\mu(1-\gamma^5)$ ($R_2$) or $1-\gamma^5$ ($R_3$). Using the equation of motion for the strange quark, and neglecting contributions of $\order{\frac{m_s}{m_b}}$,
$\hat{R}_{2,3} = \pm\frac{1}{m_b}(\bar{b}^\alpha \Gamma \gamma_0 \VEC{\gamma}\cdot \VEC{D}s^\alpha) (\bar{b}^\beta \Gamma s^\beta)$, 
and similarly for $\hat{\tilde{R}}_{2,3}$. In order to implement the derivative operator, the calculation of the associated three-point correlation functions will require new strange quark propagators, beyond those used in \cite{Dowdall:2019bea}, with point-split inversion sources at the operator location. We use a symmetric difference operator in each of the spatial dimensions as the source for the inversion of the HISQ Dirac matrix on Coulomb-gauge-fixed configurations.  Note we use a hat on an operator when we wish to call attention to the 4-quark operator computed directly on the lattice, in distinction to a linear combination such as a renormalized, matched, or subtracted operator.

As discussed below, we will also need matrix elements of the dimension-6 operators $\hat{O}_1$ and $\hat{O}_2$.  At no additional cost, we recompute these here and check that they agree with the high-statistics study \cite{Dowdall:2019bea}.

Matrix elements are obtained from fits to two- and three-point correlation functions using methods developed in Refs.~\cite{Lepage:2001ym,Hornbostel:2011hu,Bouchard:2014ypa,Dowdall:2019bea}. Some details have appeared in \cite{Davies:2017jbi}.
\iflonger
(Further details are provided in the Appendix.)
\fi

\ifsection
\section{Discussion of subtraction}
\fi

\begin{table}
  \caption{\label{tab:perturb}  
  Perturbuative coefficients used in (\ref{eq:Rsub}) for the values of $am_b$ used on each ensemble, very coarse/coarse/fine (VC/C/F) spacing with $m_s/m_l=5$ or physical (p).     }
  \centering
    \begin{ruledtabular}
  \begin{tabular}{cccccc} 
      Coeff & \multicolumn{1}{c}{VC5} & \multicolumn{1}{c}{VCp} &
    \multicolumn{1}{c}{C5} & \multicolumn{1}{c}{Cp} & \multicolumn{1}{c}{F5}
    \\ \hline
    $\xi_{21}$ & $-0.1311$ & $-0.1327$ & $-0.1557$ & $-0.1573$ & $-0.2004$ \\ 
    $\xi_{22}$ & 0.0092 & 0.0093 & 0.013 & 0.0133 & 0.0225 \\ 
    $\xi_{31}$ & $-0.0331$ & $-0.0334$ & $-0.0392$ & $-0.0397$ & $-0.0508$ \\ 
    $\xi_{32}$ & $-0.2829$ & $-0.2864$ & $-0.3404$ & $-0.3449$ & $-0.451$ 
  \end{tabular}
  \end{ruledtabular}
\end{table}

The prediction of a matrix element of a higher dimension operator using lattice NRQCD is complicated by mixing with lower dimension operators \cite{Collins:2000ix}.  The presence of the lattice cutoff $a$ means that the matrix elements $\langle\hat{R}_{2,3} \rangle$ will contain contributions from $\langle\hat{O}_{1,2}\rangle$ of the order $\order{\alpha_s/(a m_b)}$.  We have used lattice perturbation theory to calculate the coefficients $\xi_{ij}$ which parametrize this mixing at one-loop level. Matrix elements of the subtracted operator,
\begin{equation}
{R}_i^{\mathrm{sub}} = \hat{R}_i - \alpha_V  \xi_{ij} \hat{O}_j  
\label{eq:Rsub}
\end{equation}
will have power-law mixing cancelled through $\order{\alpha_s}$.  The coefficients $\xi_{ij}$ have not been calculated before.  The procedure is a straightforward extension of Ref.~\cite{Monahan:2014xra}, in particular Sec.~IV.B.  In this instance the derivative acts on the light quark propagator instead of the heavy quark propagator. Numerical values are tabulated in Table~\ref{tab:perturb}. For the numerical value of the strong coupling constant, we use $\alpha_V(2/a)$ (see Table I of \cite{Colquhoun:2015oha}, inferred from the work of \cite{McNeile:2010ji,Chakraborty:2014aca}). 

\begin{figure}
\centering
\includegraphics[width=\columnwidth]{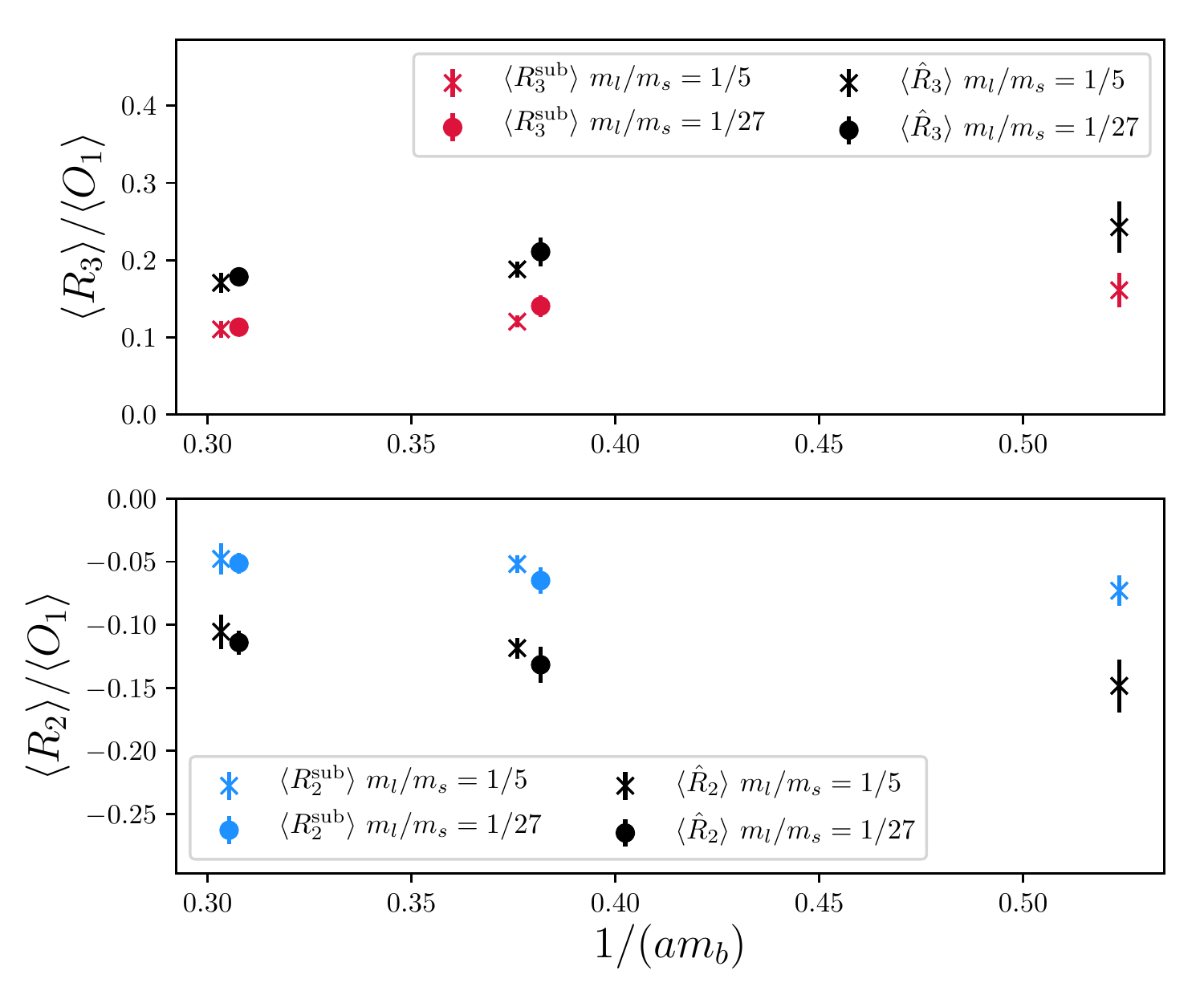}
\caption{\label{fig:ratios} Ratios of unsubtracted and subtracted $R$ matrix elements to $\langle O_1\rangle$. Only statistical errors are shown here. }
\end{figure}

Fig.~\ref{fig:ratios} illustrates the effect of the subtraction (\ref{eq:Rsub}). Matrix elements of $\hat{R}_i$ and $R_i^{\mathrm{sub}}$ are shown normalized to $\langle O_1\rangle$ \cite{Dowdall:2019bea}.  
The fact that the matrix element of the subtracted operator is 50\% ($R_2$) to 70\% ($R_3$) of the unsubtracted operator shows that this subtraction is significant.  This is of comparable size to the effects seen in the $1/m_b$ contributions to the $B$ and $B_s$ decay constants \cite{Collins:2000ix,Wingate:2003gm} and matrix elements $\langle O_j\rangle$ \cite{Dalgic:2006gp}. Fig.~\ref{fig:ratios} shows that,
as expected, the subtraction leads to a reduction in $a$-dependence over the range of lattice spacings used.

We must estimate the uncertainty due to not knowing the $\order{\alpha_s}$ $\overline{\mathrm{MS}}$-to-lattice matching of the $R$ operators nor the $\order{\alpha_s^2}$ contributions from the dimension-6 operators.  Both of these are suppressed by a power of $\alpha_s$ compared to matrix elements of the two terms in (\ref{eq:Rsub}). Therefore, we include these truncation errors in our results by multiplying our results by a noisy estimator $\langle R_i\rangle
= \langle {R}_i^{\mathrm{sub}} \rangle (1 + \alpha_V\delta_{am_b})$, 
where $\delta_{am_b} = 0\pm1$ is a Gaussian-distributed random variable, one for each of the 3 lattice spacings.  This is our largest source of uncertainty.

\ifsection
\section{Chiral-continuum fit, systematic errors}
\fi

\begin{figure}[t]
\centering
\includegraphics[width=\columnwidth]{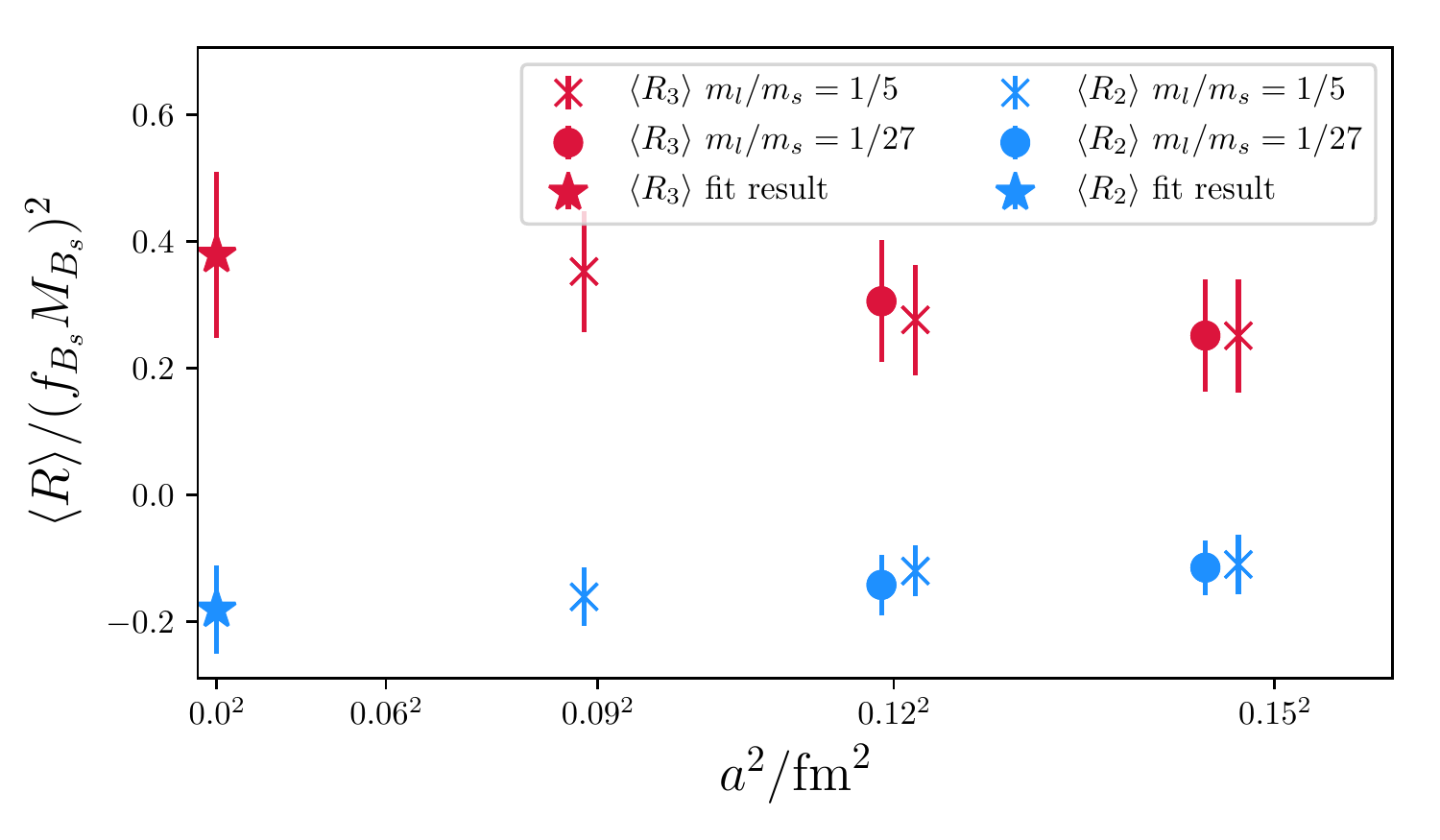}
\caption{\label{fig:r_vs_asq} Matrix elements of ${R}_2$ and ${R}_3$ plotted against lattice spacing squared. Error bars include systematic uncertainties. Results of fits to Eq.~(\ref{eq:ccfit}) are plotted as stars. }
\end{figure}

Our calculations include numerical data obtained with all quark masses tuned close to their physical values.  In order to include some data with unphysically large values for the up/down sea quark masses, we assume an analytic dependence on these masses.  We also parametrize discretization errors, e.g.\ due to the gluon and staggered fermion actions, in powers of $(a \Lambda_{\subrm{QCD}})^2$ taking $\Lambda_{\subrm{QCD}}=$ 500 MeV.  Results presented here come from fits to 
\begin{align}
\frac{\langle R_j \rangle}{f_{B_s}^2 M_{B_s}^2} = &  \beta \Big[1 + d_2 (a\Lambda_{\subrm{QCD}})^2 + d_4 (a\Lambda_{\subrm{QCD}})^4
 \nonumber \\
 & + c_1^{s,\mathrm{val}} x_s + c_1^{\mathrm{sea}}(2x_\ell' + x_s') \Big]
 \label{eq:ccfit}
\end{align}
where sea quark mass dependence is parametrized by $x_\ell' = M_\pi^2/(2\Lambda_\chi)^2$ and $x_s' = (2M_K^2 - M_\pi^2)/(2\Lambda_\chi)^2$, with $\Lambda_\chi = 1$~GeV.  We use the lattice masses $aM_\pi$ and $aM_K$ tabulated in \cite{Bazavov:2015yea}. For the valence strange quark $x_s = M_{\eta_s}^2/(2\Lambda_\chi)^2$, the $\eta_s$ being a fictitious flavor-nonsinglet $\bar{s}s$ pseudoscalar meson, which is nevertheless well-defined in chiral perturbation theory.
Values for $aM_{\eta_s}$ on these ensembles \cite{Dowdall:2013rya} are used to parametrize the difference between the input valence strange quark mass and the physical one, using for the ``physical'' value $M_{\eta_s}$ = 688.5(2.2) MeV \cite{Davies:2009tsa,Dowdall:2013rya}.  We assume Gaussian priors of $0\pm 1$ for the fit parameters, except for $d_2$, which we take to be $0\pm 5$.  In the fits, we find $|d_2| \approx 2.5\pm 2.0$.

As one would expect, any sensitivity of the $B_s - \bar{B}_s$ matrix elements to the light sea quark mass is much smaller than our uncertainties.  The slight mistunings in the sea or valence strange quark masses are not large enough to give a nonzero result for $c_1^{s}$ parameters.  Including terms quadratic in the $x$ variables has no effect on the fit. Similarly, we obtain $d_4$ consistent with zero.  Within errors the data is consistent with a mild $a^2$ dependence.

Our main results are for the pair of matrix elements 
\begin{align}
\langle B_s | R_2 | \bar{B}_s \rangle &= -(0.18 \pm 0.07) f_{B_s}^2 M_{B_s}^2 \\
\langle B_s | R_3 | \bar{B}_s \rangle &= (0.38 \pm 0.13) f_{B_s}^2 M_{B_s}^2 
\end{align}
or for the linearly dependent, color-rearranged operator matrix elements (\ref{eq:Rtilde})
\begin{align}
\langle B_s | \tilde{R}_2 | \bar{B}_s \rangle &= (0.18 \pm 0.07) f_{B_s}^2 M_{B_s}^2 \\
\langle B_s | \tilde{R}_3 | \bar{B}_s \rangle &= (0.29 \pm 0.10) f_{B_s}^2 M_{B_s}^2 \, .
\end{align}
At the accuracy with which we work, these results can be interpreted as the $\overline{\mathrm{MS}}$-scheme results at the scale $\mu_2  = m_b$, with an uncertainty included to account for the fact that the lattice-continuum matching is tree-level. 

It is sometimes convenient to include a numerical factor which arises from the vacuum saturation approximation (VSA). We define $B'$-factors via $\langle B_s | R_k| \bar{B}_s \rangle = q f_{B_s}^2 M_{B_s}^2 B'_{R_k}$  where $q$ is $-\frac23$, $\frac23$, $\frac76$, and $\frac56$ for $R_2$, $\tilde{R}_2$,  $R_3$, and $\tilde{R}_3$, respectively.
The ``unprimed'' bag factors $B_{R_k}$ which are equal to 1 in the VSA include a mass factor such that 
$B_{R_k}' = \left[ M_{B_s}^2/(m_b^{\mathrm{pow}})^2- 1\right] B_{R_k}$ \cite{Beneke:1996gn}.
These $B_{R_k}$ are the ones taken in recent phenomenological estimates \cite{Lenz:2006hd,Lenz:2011ti,Artuso:2015swg} to be $1.0\pm 0.5$ in the absence of a QCD calculation.  We tabulate numerical results of our work in Table~\ref{tab:bag}.  It turns out that the VSA expectation is a reasonable back-of-the-envelope estimate. 
(Note that while the bag factors depend on the definition of $m_b^{\mathrm{pow}}$, the $B'$-factors do not.) Our results replace the rough estimates with a lattice QCD computation with all uncertainties quantified.

\begin{table}
\centering
\caption{\label{tab:bag} Results for bag factors of operator labeled by $k$.
     The right column uses $M_{B_s} = 5.36688(17)$ GeV \cite{Tanabashi:2018oca} and $m_b^{\mathrm{pow}}=4.70(10)$ GeV
     \cite{Lenz:2006hd}.
}
\begin{ruledtabular}
\begin{tabular}{cccccc}
$k$ & \multicolumn{1}{c}{$B_{R_k}'$} & \multicolumn{1}{c}{$B_{R_k}$} & $k$ & \multicolumn{1}{c}{$B_{R_k}'$} & \multicolumn{1}{c}{$B_{R_k}$}\\ \hline \\[-3mm]
$R_2$ & 0.27(10) & 0.89(38) & 
$\tilde{R}_2$ & 0.27(10) & 0.89(38) \\ 
$R_3$ & 0.33(11) & 1.07(42) & 
$\tilde{R}_3$ & 0.35(13) & 1.14(46) \\
\end{tabular}
\end{ruledtabular}
\end{table}

The matrix elements determined in Ref.~\cite{Dowdall:2019bea} allow determination of the remaining 3 matrix elements
$B_{R_0}' = -\frac34 {\langle B_s | R_0 | \bar{B}_s \rangle}/(f_{B_s}M_{B_s})^2 = 0.32(13)$, 
$B_{R_1}' = \frac37 {\langle B_s | O_4 | \bar{B}_s \rangle}/(f_{B_s}M_{B_s})^2 =1.564(64)$, and
$B_{\tilde{R}_1}' = \frac35 {\langle B_s | O_5 | \bar{B}_s \rangle}/(f_{B_s}M_{B_s})^2 = 1.167(46)$.

\ifsection
\section{${\Delta\Gamma_s}$ in the Standard Model}
\fi

Our results permit the first lattice determination of $\Delta\Gamma_{1/m_b}$, the power-law corrections to $\Delta\Gamma_s$.  Recently there has been an investigation of scale and scheme dependence of the leading term in $\Delta\Gamma_s$, where it has been proposed to include corresponding uncertainties as follows \cite{Asatrian:2017qaz}
\begin{equation}
\Delta \Gamma_s = [1.86(17) B_1 + 0.42(3) B_3'] f_{B_s}^2 + \Delta\Gamma_{1/m_b}
\label{eq:DGs_LO}
\end{equation}
in the $\overline{\mathrm{MS}}$ scheme.  Taking $f_{B_s} = 0.2307(12)$ GeV from Ref.~\cite{Bazavov:2017lyh} and weighted averages of $B_1=0.84(3)$ and $B_3'=1.36(8)$ from Refs.~\cite{Bazavov:2016nty,Dowdall:2019bea} yields a result for the leading order contribution, $\Delta \Gamma_s^{\mathrm{LO}} = 0.114(9)$ ps${}^{-1}$.

The $1/m_b$ contribution to $\Delta \Gamma_s$ can be expressed as a linear combination of the matrix elements of the $R$ operators, times perturbative coefficients $\gamma_k$ \cite{Beneke:1996gn,Lenz:2006hd}.  Writing $\Delta \Gamma_{1/m_b} = -2\tilde\Gamma_{12,1/m_b} \cos \phi_{12}$, we have
\begin{equation}
 \tilde{\Gamma}_{12,1/m_b} = \frac{G_F^2 f_{B_s}^2 M_{B_s}m_b^2}{24 \pi} \sum_{k} {\gamma}_k(\bar{z})  B_{{R}_k}' \,.
 \label{eq:DGs_NLO}
\end{equation}
Here $k$ is an index that runs over the 4 operators in (\ref{eq:Rops}) plus the 3 color-rearranged operators.
The $\gamma_k(\bar{z})$ are related to the $g_k(\bar{z})$ of \cite{Lenz:2006hd} by the numerical coefficients relating the matrix elements to the $B'$ factors; additionally $\gamma_1$ and $\tilde\gamma_1$ include a factor of $\bar{m}_s(\bar{m}_b)/\bar{m}_b(\bar{m}_b)$. The functions $g_k(\bar{z})$ depend on $\bar{z} = (\bar{m}_c(\bar{m}_b)/\bar{m}_b(\bar{m}_b))^2$ \cite{Beneke:2002rj} and the leading order $H^{\Delta F = 1}$ Wilson coefficients $C_1^{(0)}$ and $C_2^{(0)}$.  
\iflonger
Numerical values used here are given in the Appendix.  
\fi
Only the terms with $B_{R_0}'$, $B_{R_2}'$, and $B_{\tilde{R}_2}'$ contribute to $\Delta\Gamma_{1/m_b}$ due to the smallness of the other $\gamma_k$. 

Our result for (\ref{eq:DGs_NLO}) is $\tilde\Gamma_{12,1/m_b} =  0.0110(52) ~\mathrm{ps}^{-1}$ which, given $\cos\phi_{12} = 1$ to the precision relevant here, contributes to the width difference as
\begin{equation}
\Delta \Gamma_{1/m_b} = -2 \tilde\Gamma_{12,1/m_b} = -0.022(10) ~\mathrm{ps}^{-1} \,.
\label{eq:DGcorr}
\end{equation}
The uncertainty in (\ref{eq:DGcorr}) is dominated by that of $\langle \tilde{R}_2\rangle$.
From studies of the leading $\Delta \Gamma_s$ term, we expect scale and scheme uncertainties here to similarly be at the 10\% level, i.e.\ not significant compared to the present hadronic uncertainty.

Combining (\ref{eq:DGcorr}) with (\ref{eq:DGs_LO}), we find 
\begin{equation}
\Delta\Gamma_s = 0.092(14)~\mathrm{ps}^{-1} \,.
\label{eq:DGs}
\end{equation}
The error in (\ref{eq:DGs}) is mostly due to the uncertainty in $\Delta \Gamma_{1/m_b}$; its variance contributes approximately 60\% to the total variance in (\ref{eq:DGs}).  The next largest uncertainty, contributing 30\%, comes from the coefficients in first term of (\ref{eq:DGs_LO}).  The variance of $B_1$ contributes $8\%$ of the total.

\ifsection
\section{Conclusions}
\fi

The HFLAV average of experimental measurements is $\Delta\Gamma_s = 0.085(6)$ ps${}^{-1}$ \cite{Amhis:2019aaa}.  This is in good agreement with our result (\ref{eq:DGs}) with half the uncertainty.  

There remains more to do in order for the theoretical prediction to match the experimental precision.  The next generation lattice calculation will require one-loop matching of lattice to $\overline{\mathrm{MS}}$ regularization schemes in order to reduce the uncertainty in $\Delta\Gamma_{1/m_b}$.  At the same time the work to determine the perturbative coefficients appearing in (\ref{eq:DGs_LO}) through NNLO must be completed. First steps have already begun \cite{Asatrian:2017qaz}.


\textit{Acknowledgments:} 
We thank the MILC collaboration for their gauge configurations and
their code MILC-7.7.11 \cite{MILCgithub}.  
MW is grateful for an IPPP Associateship held while some of this work was undertaken and for
discussions with A.~Lenz. This work was funded in part by the STFC, the NSF, and the DOE.
CJM is supported in part by the U.S.\ Department of Energy, Office of Science, Office of Nuclear Physics under contract Nos.\ DE-FG02-00ER41132 and DE-AC05-06OR23177.
Results described here were obtained using the
Darwin Supercomputer of the University of Cambridge High Performance
Computing Service as part of the DiRAC facility jointly funded by
STFC, the Large Facilities Capital Fund of BIS and the Universities of
Cambridge and Glasgow. 

\bibliographystyle{h-physrev5_collab.bst}
\bibliography{mbw}


\iflonger

\clearpage
\appendix
\section*{APPENDIX}

In this Appendix we collect a few pertinent details from the literature cited in the main body, and we give a few numerical values for the short-distance coefficients entering the Standard Model calculation of $\Delta \Gamma_{1/m_b}$.

\begin{table}
\caption{\label{tab:configs}Parameters of the MILC $n_f=2+1+1$ HISQ configurations used.
Masses listed are sea quark masses.  Lattice spacing determined using the $\Upsilon$
splittings, as in Table I of \cite{Dowdall:2013tga}; errors are statistical,
NRQCD systematic, experiment respectively. }
\begin{ruledtabular}
\centering
\begin{tabular}{cllllcr}
Label & \multicolumn{1}{c}{$a$/fm} & 
\multicolumn{1}{c}{$am_l$} & \multicolumn{1}{c}{$am_s$} & 
\multicolumn{1}{c}{$am_c$} & $N_s^3\times N_t$ & \multicolumn{1}{c}{\#} \\
\hline
VC5 &  0.1474(5)(14)(2) & 0.013 & 0.0650 & 0.838 & $16^3\times 48$ & 1020 \\
VCp &  0.1450(3)(14)(2) & 0.00235 & 0.0647 & 0.831 & $32^3\times 48$ & 1000 \\
C5 & 0.1219(2)(9)(2) & 0.0102 & 0.0509 & 0.635 & $24^3\times 64$ & 1052 \\
Cp & 0.1189(2)(9)(2) & 0.00184 & 0.0507 & 0.628 & $48^3\times 64$ & 1000 \\
F5 & 0.0873(2)(5)(1) & 0.0074 & 0.037 & 0.440 & $32^3\times 96$ & 1008 
\end{tabular}
\end{ruledtabular}
\end{table}

\begin{table}
\caption{\label{tab:valence}Valence quark parameters, with $c_2 =c_3=1$. }
\centering
\begin{ruledtabular}
\begin{tabular}{clllccc}   
Ensemble & \multicolumn{1}{c}{$am_s^{\mathrm{val}}$} & 
\multicolumn{1}{c}{$am_b$} & \multicolumn{1}{c}{$u_{0L}$} &
$c_1=c_6$ & $c_4$ & $c_5$  \\ \hline
VC5 & 0.0641 & 3.297 & 0.8195  & 1.36 & 1.22 & 1.21 \\
VCp & 0.0628 & 3.25 & 0.819467 & 1.36 & 1.22 & 1.21  \\
C5 & 0.0522 & 2.66 & 0.834 & 1.31 & 1.20 & 1.16 \\
Cp & 0.0507 & 2.62 & 0.834083 & 1.31 & 1.20 & 1.16  \\
F5 & 0.0364 & 1.91 & 0.8525 & 1.21 & 1.16 & 1.12  
\end{tabular}
\end{ruledtabular}
\end{table}

\begin{table}[t]
\centering
\caption{\label{tab:HQEcoefficients} Numerical values used in this Standard Model prediction of $\Delta \Gamma_{1/m_b}$. (See text for discussion and references.) The quark masses and Wilson coefficients determine the $\gamma$ coefficients (using $\bar{z}$). Errors in these tabulated values are much smaller than in the matrix elements they multiply.}
\begin{ruledtabular}
\begin{tabular}{cd}
quantity &\multicolumn{1}{c}{\textrm{value}}  \\ \hline
$\bar{m}_c(\bar{m}_c)$/GeV & 1.2753(65)  \\
$\bar{m}_c(\bar{m}_c)/\bar{m}_c(\bar{m}_b)$ & 1.392(9) \\
$\bar{m}_b(\bar{m}_b)$/GeV &  4.195(14)    \\
$C_1^{(0)}$ &  -0.269 \\
$C_2^{(0)}$ & 1.12 \\ 
$\gamma_0$ & 	  	 0.505 \\
$\gamma_1$ & 	 	 0.033 \\
$\tilde\gamma_1$ & 	  -0.077 \\
$\gamma_2$ & 	  	 -0.513 \\
$\tilde\gamma_2$ & 	 -1.667 \\
$\gamma_3$ & 		 0.027 \\
$\tilde\gamma_3$ & 	 -0.063 \\
\end{tabular}
\end{ruledtabular}
\end{table}

The parameters of the MILC gauge field ensembles we used are given in Table~\ref{tab:configs} \cite{Bazavov:2010ru,Bazavov:2012xda,Bazavov:2015yea}, and the valence quark parameters we used in Table~\ref{tab:valence} \cite{Dowdall:2011wh,Dowdall:2013tga,Dowdall:2019bea}.

In order to determine the energies and amplitudes associated with the meson creation and annihilation operators, we perform multi-exponential Bayesian fits \cite{Lepage:2001ym}.  The fit functions for the two-point and three-point functions are, respectively,
\begin{align}
  C^{\mathrm{2pt}}_{ab}(t) & = \sum_{i=0}^{N_\mathrm{2pt}-1} X_{a,i} X_{b,i} e^{-E_i t}
  - (-1)^{t/a} Y_{a,i} Y_{b,i} e^{-E_i^o t}
  \label{eq:C2pt}
  \intertext{and, abbreviating terms containing any $Y$ parameters,}
  C^{\mathrm{3pt}}_{ab}(t,T) & = \sum_{i,j=0}^{N_{\mathrm{3pt}}-1} X_{a,i} V_{nn,ij} X_{b,j}
  e^{-E_i t} e^{-E_j(T-t)} +\mathrm{osc.}
  \label{eq:C3pt}
\end{align}
In practice, we fit to energy differences for all but the ground state energy $E_0$.  The $X$ and $Y$ parameters are the amplitudes for meson creation/annihilation. The labels $a$ and $b$ run over the 3 types of smearings used with the $B_s$ meson interpolating operators, a local operator and two Gaussian-smeared operators with different widths \cite{Colquhoun:2015oha}.  The $V_{nn,ij}$ are parameters related to the matrix elements of 4-quark operators. Not all fit parameters are well-constrained by the data, so we introduce Bayesian priors with means and widths chosen as described in Ref.~\cite{Davies:2017jbi}.

We take the two-point functions from earlier studies of $B_{(s)}$ decay constants \cite{Dowdall:2013tga,Colquhoun:2015oha}; these were obtained with 16 time sources on each gauge-field configuration, allowing precise determination of the ground state energies and decay amplitudes. Given that our determinations of the matrix element of $R_{2,3}$ will have $\order{\alpha_s}$ truncation errors, we do not need such high statistics.  The three-point correlators in this work come from using just 2 time sources on each configuration.  

We find the best approach in this case is to use chained \cite{Bouchard:2014ypa}, marginalized fits \cite{Hornbostel:2011hu}.  The high-statistics two-point functions are fit using $N_{\mathrm{2pt}}=5$.  The resulting $E_I$ and $X_{a,i}$, central values and errors, are used as priors for the fits to the three-point functions.  Excited state contamination is accounted for in the three-point functions by incorporating noisy estimates using the results of the two-point fits, then fitting using $N_{\mathrm{3pt}}=1$ \cite{Hornbostel:2011hu,Dowdall:2019bea}. 

While we perform separate fits to the correlation functions associated with each $\hat{R}_{2,3}$ and $\hat{O}_i$ in order to study the relative contributions to the subtracted matrix elements, for the main results we perform fits to linear combinations of three-point functions so that the subtracted matrix element is directly extracted from the fit.  This allows correlations to be propagated straightforwardly.
 
Table~\ref{tab:HQEcoefficients} gives numerical values for the short-distance quantities used in combination with our matrix elements to determine $\Delta\Gamma_{1/m_b}$. Numerical values for the Wilson coefficients have been taken from Table 2 of Ref.~\cite{Asatrian:2017qaz}. 
For the charm quark mass we use the world average \cite{Lytle:2018evc} of lattice results with $2+1+1$ flavors of sea quarks \cite{Carrasco:2014cwa,Chakraborty:2014aca,Bazavov:2018omf,Lytle:2018evc}, and for the bottom mass and the ratio $m_b/m_c$ we use the result from Ref.~\cite{Bazavov:2018omf}. 

\fi 


\end{document}